\documentclass[conference]{IEEEtran}
\IEEEoverridecommandlockouts
\usepackage{cite}
\usepackage{amsmath,amssymb,amsfonts}
\usepackage{algorithmic}
\usepackage{graphicx}
\usepackage{subfigure}
\usepackage{textcomp}
\usepackage{xcolor}
\usepackage{booktabs}
\usepackage{multirow}
\usepackage{caption}
\makeatletter
\newcommand{\linebreakand}{%
  \end{@IEEEauthorhalign}
  \hfill\mbox{}\par
  \mbox{}\hfill\begin{@IEEEauthorhalign}
}
\makeatother
\def\BibTeX{{\rm B\kern-.05em{\sc i\kern-.025em b}\kern-.08em
    T\kern-.1667em\lower.7ex\hbox{E}\kern-.125emX}}
\begin{document}

\title{MaxMind: A Memory Loop Network to Enhance Software Productivity Based on LLMs\\

}

\author{\IEEEauthorblockN{1\textsuperscript{st} Yuchen Dong}
\IEEEauthorblockA{\textit{Independent Researcher} \\
\textit{ Hunan Province}\\
Changsha, China \\
yuchenworks@yeah.net}
\and
\IEEEauthorblockN{2\textsuperscript{nd} Xiaoxiang Fang}
\IEEEauthorblockA{\textit{School of Computer} \\
\textit{ National University of Defence Technology}\\
Changsha, China \\
xiaoxiang200@163.com}
\linebreakand 
\IEEEauthorblockN{3\textsuperscript{rd} Yuchen Hu}
\IEEEauthorblockA{\textit{School of Integrated Circuits} \\
\textit{ Southeast University}\\
Nanjing, China \\
yuchen\_hu@seu.edu.cn}
\and
\IEEEauthorblockN{4\textsuperscript{th} Renshuang Jiang}
\IEEEauthorblockA{\textit{School of Computer} \\
\textit{ National University of Defence Technology}\\
Changsha, China \\
renshuang717@163.com}
\and
\IEEEauthorblockN{5\textsuperscript{th} Zhe Jiang}
\IEEEauthorblockA{\textit{School of Integrated Circuits} \\
\textit{ Southeast University}\\
Nanjing, China \\
zhejiang.uk@gmail.com}

}

\maketitle

\begin{abstract}
The application of large language models to facilitate automated software operations and tool generation (SOTG), thus augmenting software productivity, mirrors the early stages of human evolution when the ability to create and use tools accelerated the progress of civilization. 
These complex tasks require AI to continuously summarize and improve. 
Current research often overlooks the importance of converting real-time task experiences into system memory and differentiating the value of existing knowledge for future reference. 
This paper addresses these issues by evolving external memory models into Memory-Loop Networks for timely memorization and experience referencing. 
We also enhance a RAG mechanism with knowledge precision segmentation to utilize memory based on value differentiation, and design the MaxMind model for SOTG accordingly.
To demonstrate our approach, we developed MaxMind4Sheet, an electronic spreadsheet processing system aligned with the MaxMind philosophy. 
Comparative experiments with SheetCopilot have demonstrated that the accumulation and recycling of task memories lead to a steady enhancement in task success rate, with an improvement rate of approximately 3\%-6\% per round in this implementation example. Note that as the memories continue to grow, this cumulative improvement may be substantial. The inclusion of memory recycling can also boost the system's task execution efficiency by up to 25\%, and it can address the retraining issue faced by LLMs when handling specialized tasks through memories transfer.
These suggest that MaxMind has significant potential to enhance the capabilities and productivity of LLM systems in SOTG.

\end{abstract}


\section{Introduction}
The rapid evolution of Large Language Models (LLMs) has spotlighted their potential in facilitating Software Operations and Tool Generation (SOTG), leveraging their advanced natural language understanding and reasoning capabilities~\cite{wei2022chain}~\cite{shen2024hugginggpt}~\cite{bran2023chemcrow}. 
A typical example is SheetCopilot~\cite{li2024sheetcopilot}, which leverages a curated suite of atomic actions to automate spreadsheet program creation, achieving a commendable success rate. 
In contrast to typical Q\&A(question answering) system\cite{Allemang2024IncreasingTL}\cite{surveyQA}, SOTG endeavors are inherently more intricate, necessitating a heightened level of professional proficiency and rigorous reasoning skills.

\begin{figure}[htb]
\centering
\includegraphics[width=0.45\textwidth, trim=0cm 3cm 11cm 0cm, clip]{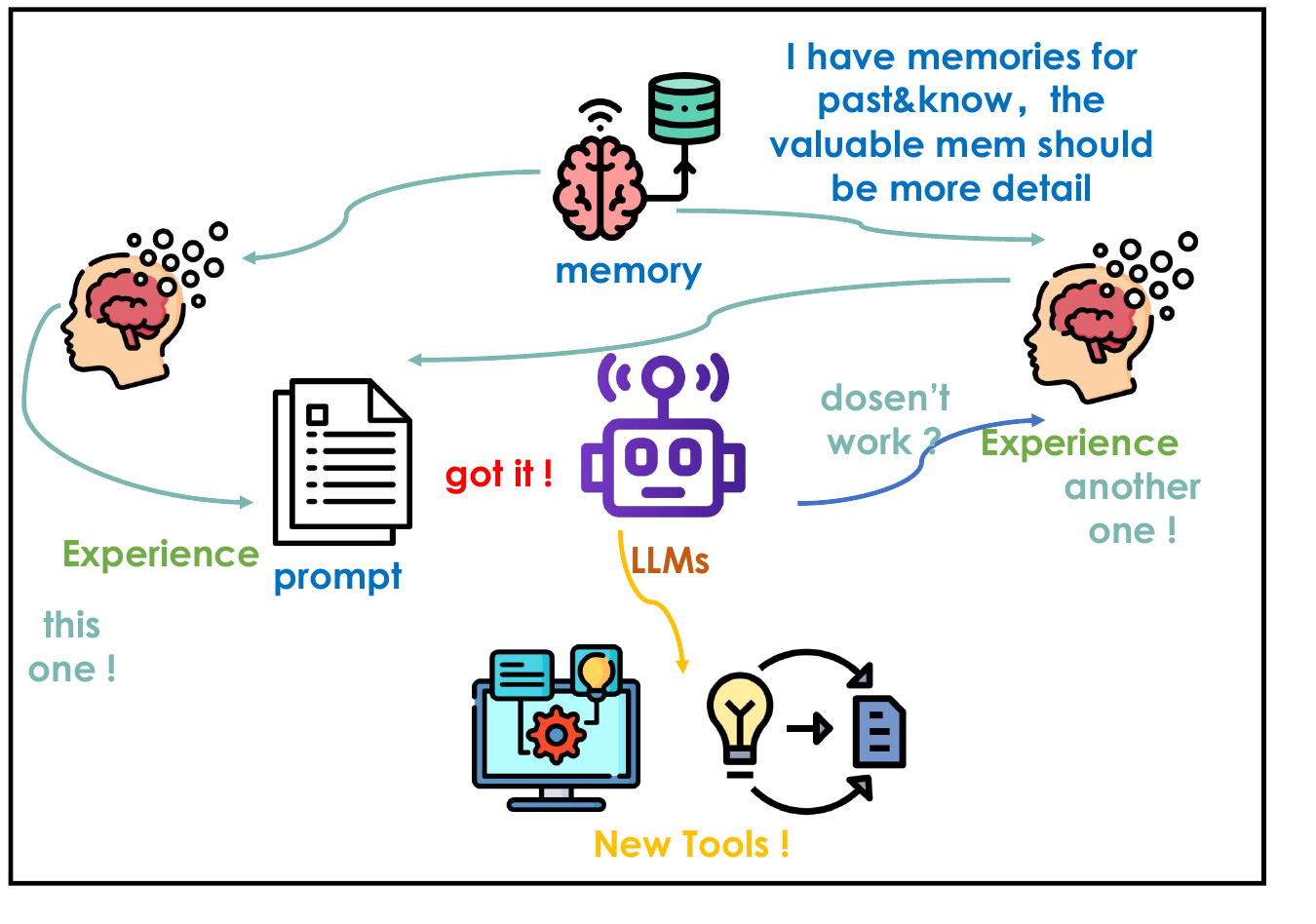}
\caption{Our basic idea: the past experience and the different precise of memory can be helpful to future work}
\vspace{-1.0em}
\label{fig0}
\end{figure}

Current approaches, grounded in Chain of Thought (CoT)~\cite{wei2022chain}~\cite{yang2023mm} or Retrieval-Augmented Generation (RAG)~\cite{mialon2023augmented}~\cite{jiang2024longrag}, frequently integrate domain-specific knowledge repositories. 
However, these methods encounter a significant performance decline when confronted with highly intricate or ambiguous tasks, owing to the limitations in timely updating and enhancing the knowledge base~\cite{hao2022structured}~\cite{ratner2022parallel}. This predicament leads to an escalation in error rates and failures, underscoring the need for more dynamic and adaptable solutions. Moreover, the existing frameworks exhibit limited retraining capabilities, with their performance faltering once they venture beyond the confines of the current dialogue context. 
The absence of a mechanism to retain and leverage past experiences and lessons learned undermines their potential for continuous improvement~\cite{li2024flexkbqa}~\cite{yang2023beyond}. 
Consequently, empowering the system with learning abilities that facilitate experience accumulation and enhanced task reasoning is paramount to advancing SOTG tasks.
This necessitates the development of innovative strategies that can dynamically adapt, learn from mistakes, and continually refine their performance in the face of ever-evolving challenges (as shown in Fig \ref{fig0}).

To revolutionize software productivity in the context of LLMs, this paper introduces a new Memory-Loop Network (MLN) idea, embodied in our novel  MaxMind system model. 
At its core, MaxMind leverages a memory recycling mechanism to dynamically and timely update the external knowledge library. It implements a novel approach that incorporates knowledge relevance and task execution feedback into RAG, enhancing its capabilities. 
This system instantaneously archives inferences and task executions in an external memory repository. When confronted with a new task, MaxMind formulates the task demands into a query, retrieves highly pertinent memories (knowledge), and automatically adjusts precision to ensure better performance.

The contributions of this paper are summarized as follows:
\begin{enumerate}
    \item Novel Memory-Loop Network (MLN): Unlike contemporary Memory Network paradigms, MLN's output serves as a reference base for subsequent request inputs, enabling a transformative feedback loop where outputs are reintroduced as inputs, dynamically updating the memory slots. This ascending cycle facilitates the system's capacity to assimilate knowledge from operational experiences, enrich its memory bank, and iteratively refine the quality and prowess of subsequent services.

    \item Adaptive RAG Methodology Driven by Knowledge Value: We propose an innovative RAG approach grounded in knowledge value assessment. This methodology assigns graduated levels of precision to identical knowledge entities, dynamically selecting the appropriate precision level based on the task's relevance. Furthermore, it addresses token size constraints through synchronous precision scaling, optimizing RAG's efficiency both in terms of knowledge volume and precision.


    \item Prototype Realization and Experimental Validation: building upon the llama3.1-70B model and a vector database, we implement a prototype system, MaxMind4Sheet, embodying the aforementioned advancements. A series of meticulously designed experiments were conducted to gauge the system's performance and scalability, conclusively demonstrating the efficacy and practicality of the introduced techniques.
\end{enumerate}

\begin{figure*}[htp]
\centering
\subfigure[Memory network]{
\begin{minipage}[t]{0.45\textwidth}
\label{fig:memnn}
\includegraphics[width=1.8\textwidth, trim=0cm 11cm 2cm 0cm, clip]{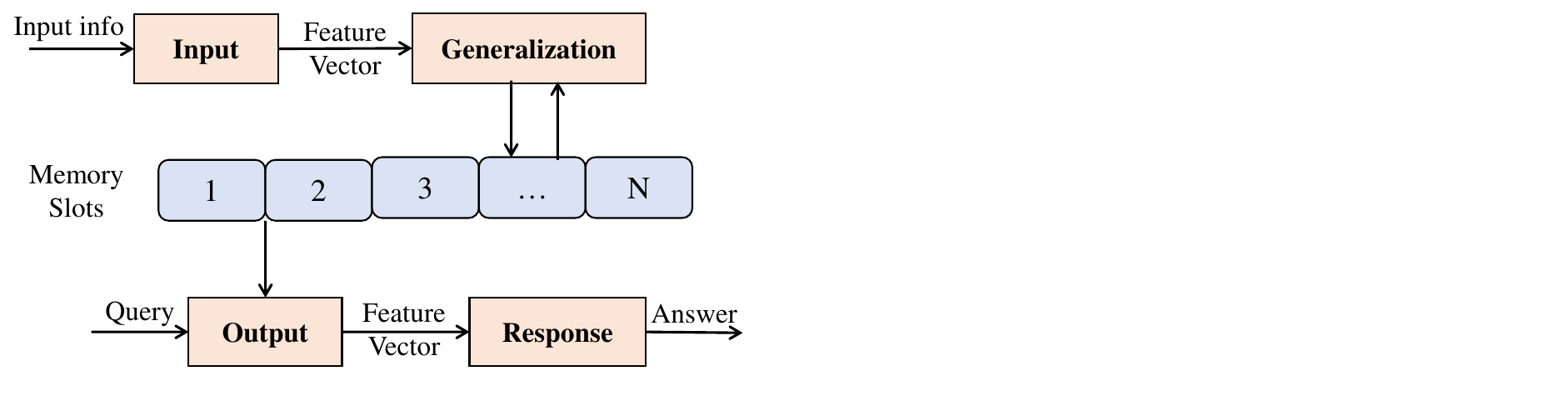}
\end{minipage}
}
\subfigure[Memory loop network]{
\begin{minipage}[t]{0.45\textwidth}
\label{fig:mln}
\includegraphics[width=1.7\textwidth, trim=0cm 10.5cm 2cm 0cm, clip]{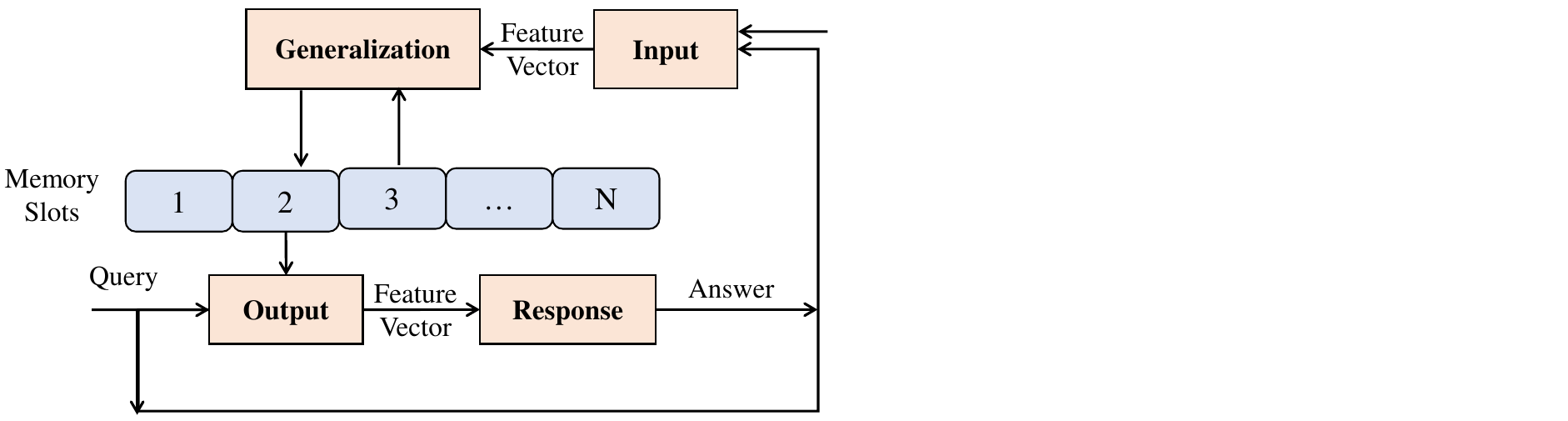}
\end{minipage}
}
\caption{From Memory Network to Memory Loop Network}
\vspace{-1.0em}
\label{fig1}
\end{figure*}

\section{Related Works}

\subsection{LLM-Aided Software Operation and Tool Generation}
Researchers capitalize on the capabilities of LLM in comprehension and transactional reasoning  to integrate it seamlessly with complementary tools and implementation methodologies, ultimately fostering a remarkable enhancement in productivity. Paranjape et al.~\cite{paranjape2023art} automates multi-step reasoning and tool selection, reducing manual programming needs. WebGPT~\cite{nakano2021webgpt} revolutionizes Web browsing with text-centric refinement, enhancing retrieval and generation via imitation and reinforcement learning. SheetCopilot~\cite{li2024sheetcopilot} demonstrates LLM-driven closed-loop control in spreadsheets, generating tailored solutions for efficient automation. These research leverage external tools to bridge LLM gaps, enhancing reasoning and task execution. This synergy strengthens AI's real-world comprehension and introduces novel solutions for challenges like semantic comprehension with fundamental functions. However, current advancements focus on LLMs' self-improvement and tool integration, primarily showcasing their capacity to learn and harness tools. This limits task diversity and also limits success rates for intricate tasks.
\subsection{Retrieval-Augmented Generation}
RAG empowers LLMs to extract info from diverse sources, boosting precision and reducing hallucinations. Recent focus include:
(1) Retrieval Enhancement: Borgeaud et al.\cite{borgeaud2022improving} integrates retrieval-augmented training, optimizing parameters and reducing complexity. ANCE \cite{xiong2020approximate} constructs realistic negative samples via ANN index updates. Karpukhin et al.\cite{karpukhin2020dense} uses contrastive learning and similarity matrix for efficient answer extraction. However, challenges remain in capturing complex semantics and overlooking info, necessitating further RAG innovations.
(2) Boosting Generator: cheng et al.\cite{cheng2024lift} fine-tunes generator via memory pool, minimizing KL loss for precise memory selection. Asai et al.\cite{asai2019learning} improves text quality and factuality through retrieval and self-reflection. Current methods face memory capacity limits, hindering wide info coverage and adaptation to changing needs.
(3) Integrating \& Fine-tuning: Shi et al. \cite{shi2023replug} enhance LM with adjustable retrievers. Cheng et al. \cite{cheng2021unitedqa} create versatile retrievers for zero-shot tasks. lin et al. \cite{lin2023ra} and jiang et al. \cite{jiang2024longrag} optimize generator and retriever for better recall. However, black-box LLMs limit interpretability and flexibility, necessitating long-embedding models.
In summary, RAG confronts key challenges: (1) Index accuracy. Content chunking and size determination may result in omissions or noise in answers. (2) Retrieval limitations. The query process occasionally missing relevant documents, extracting irrelevant context, or providing incomplete answers, hindering the full utilization of potential information. These issues impose stringent constraints on SOTG tasks.
\subsection{Integration of Neural Networks with External Memory}
Neural networks often struggle with long-term memory, hindering their ability to grasp associations over vast information spans and limiting applicability in contexts demanding knowledge accumulation. Memory Networks (MemNN)~\cite{weston2014memory}, featuring a dedicated memory module to store diverse information. By correlating inputs with memory content, MemNN extracts relevant data to generate outputs, fusing machine learning with deductive power through a read-write memory, fostering a dynamic long-term knowledge base while honing reasoning skills.
Expanding on MemNN, Sainbayar et al.\cite{sukhbaatar2015end} devised an end-to-end training approach that iteratively mines valuable information for multi-faceted reasoning. Kumar et al.\cite{kumar2016ask} advanced MemNN to handle dynamic linguistic inputs and complex relationships, introducing DMN. 
As an alternative approach to augmenting neural networks with memory capabilities, Google propose Neural Turing Machine (NTM)~\cite{graves2014neural} in its DeepMind. NTM comprises a neural controller and memory module. The controller interacts with the environment via input/output vectors, managing memory read/write through a selective matrix. Key to its trainability, NTM's differentiable components facilitate gradient descent. MemNN and NTM both have their own memory, input, read, write, and output modules. However, MemNN's memory leans toward read optimization, while NTM's write module, crucial for capacity constraints, distinguishes it by deciding rewrite/erase zones. 
While enhancing task-specific performance, these methods have limitations. They rely on costly manual annotations and positively evaluated data, limiting the model's generalization to new, unseen data. Exclusive reliance on positive feedback may restrict its ability to identify and rectify negative attributes, hindering broad applicability.

\section{Memory-Loop Network and MaxMind Model}
\subsection{Evolution of MemNN:  Memory-Loop Network}\label{AA}
The cornerstone of executing intricate SOTG resides in addressing two pivotal challenges: firstly, fostering the accumulation of experience and ensuring the agility of knowledge bases through flexible, on-demand updates; secondly, integrating an ample quantity of external knowledge while judiciously assessing the value disparities among diverse knowledge sources, e.g., allocating token quotas proportionately to their significance within the confines of token limitations.

It is widely acknowledged that the pinnacle achievements of human intelligence lie in our capacity to autonomously wield and innovate tools, distill valuable lessons from both personal and collective past experiences, and promptly harness these insights to refine subsequent actions. These defining traits serve as indispensable benchmarks for empowering AI to execute software tasks and devise tools. Consequently, this paper, through a comprehensive juxtaposition of extensive AI models against human intelligence, presents a key observation alongside three profound insights.

\textbf{Observation}: LLMs, owing to their substantial training expenses, become static post-training, primarily serving as inference engines. The diminishing of their memory capabilities emerges as a main constraint, hindering the expansion of system functionalities and overall performance~\cite{brown2020language}~\cite{roberts2020much}.

\textbf{Insight 1}: The evolution of human intelligence is underpinned by relentless learning and practice, fostering a virtuous cycle that intertwines memory, reasoning, and action. This continuous cycle fosters the gradual enhancement and adaptation of our capabilities in real-world contexts.

\textbf{Insight 2}: In the realm of reasoning, humans rely heavily on the relevance of their stored memories to the problem at hand. This relevance serves as a main basis, guiding us to evaluate and prioritize our memories, often necessitating strategic precision trade-offs.

\textbf{Insight 3}: As humans embark on tasks, we dynamically adjust our plans in response to the consequences of our actions. This adaptability not only shapes our future actions but also influences the application of past memories, subtly shifting the weight of each experience in shaping our decision-making and planning processes.

Drawing from the Observation, integrating memory components with LLMs is imperative for bolstering the functionality and broadening the capabilities of intelligent systems. This synergy echoes the principles of MemNN models, necessitating the integration of memory input, updating, output, and response functionalities. Guided by Insight 1, by linking the response feedback within the MemNN back to its input, we can establish a cyclic memory network, enabling subsequent memory outputs to dynamically harness and refine previous experiences, thereby progressively elevating the quality of outputs.
Incorporating Insight 2, we introduce a relevance assessment at the memory output stage, strategically aligning the relevance and quality of memory output and response, ensuring that more pertinent and high-quality information contributes to reasoning.
Building upon Insight 3, we devise a mechanism where memory responses retroactively influence memory outputs, fostering an iterative cycle of improvement in task reasoning and execution. This approach ensures that as tasks unfold, the system continually refines its memory usage and response strategies.

\begin{figure*}[htb]
\centering
\includegraphics[width=0.87\textwidth]{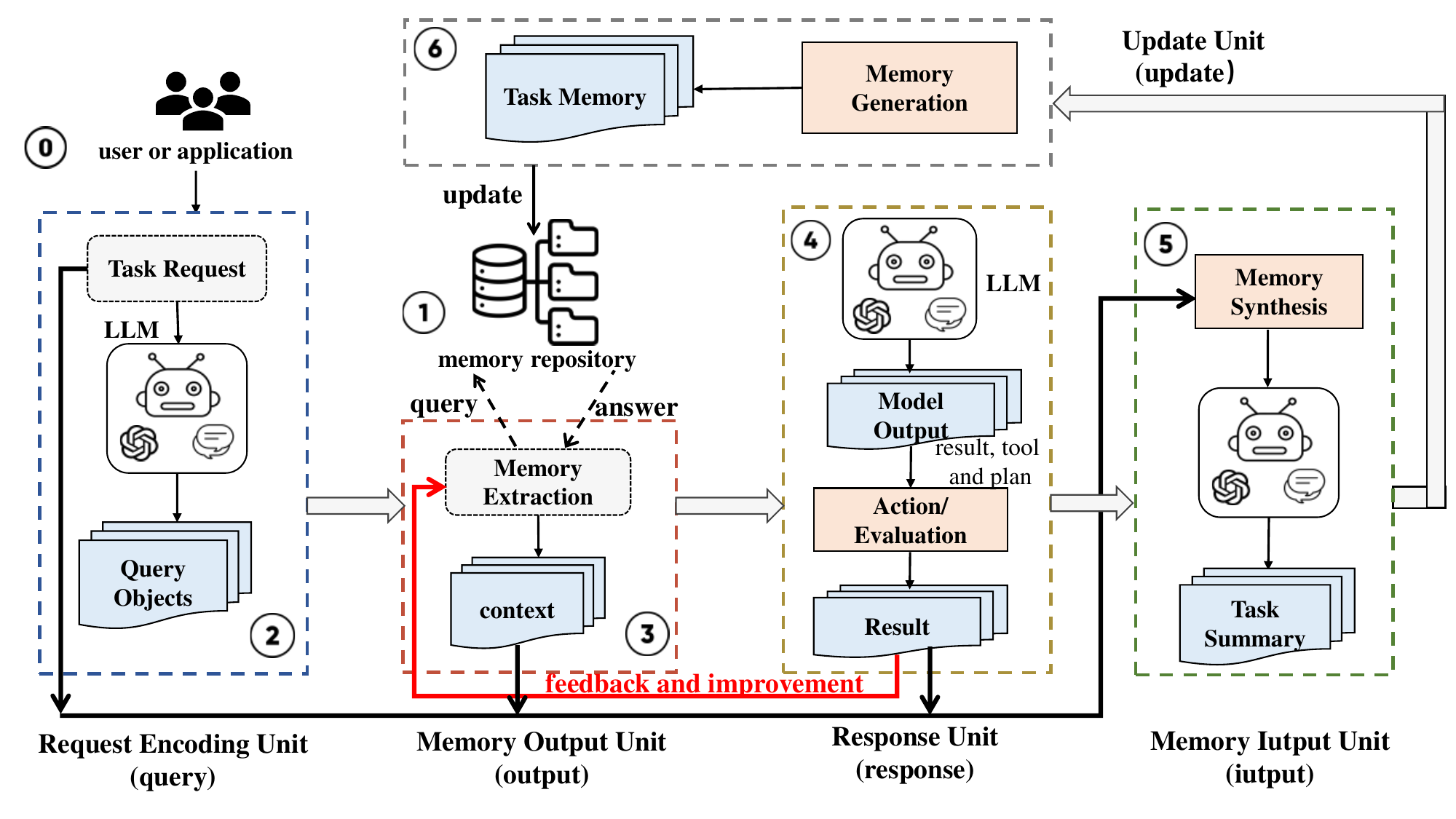}
\caption{MaxMind Model}
\vspace{-1.0em}
\label{fig2}
\end{figure*}

Drawing upon these insights, this paper delves into enhancing the classic Memory Network model, evolving it into a Memory-Loop Network(MLN), as illustrated in the Fig \ref{fig1}. The conventional Memory Network model (Fig \ref{fig:memnn}) comprises five distinct modules: memory slots, an input module, an update module, an output module, and a response module. In its operation, memory input updates and memory responses to queries function as separate processes, with the latter having no direct influence on the former. Notably, the traditional model necessitates a dedicated channel for memory updates. The most salient improvement of the MLN (Fig \ref{fig:mln}) lies in its capability to reintroduce the network's outputs, responses, and other contents as input information, thereby enabling the network to update its memory while responding to external queries. Consequently, in subsequent tasks, the network is endowed with the ability to refer back to previous task processes, refining its performance. This gives rise to a virtuous cycle of capability enhancement.

Within the realm of the MLN structure, we delve deeper by integrating the relevance filtering of memories, as inspired by Insight 2, and harnessing LLMs within the memory screening, input, output, and response processes. Furthermore, we incorporate the response feedback mechanism outlined in Insight 3, culminating in the foundational model of MaxMind. This integrated approach ensures that MaxMind is equipped to not only sieve through memories for relevance but also dynamically adjust its memory utilization based on feedback, thereby refining its reasoning and performance over time.

\subsection{MaxMind Model}\label{AA}

The design of the MaxMind model is meticulously crafted to explore the full potential of LLMs, aiming to elevate the performance of MLN structure to new heights. Therefore, each component of model has undergone a unique and thoughtful redesign, as elegantly portrayed in Figure \ref{fig2}. This approach ensures that every facet of the model synergistically contributes to its overall effectiveness, capitalizing on the strengths of LLMs and driving innovative advancements in MLN.

MaxMind's storage component can store large amount of memories, and we term it the "Memory Repository (MR)". MR supports diverse implementation methodologies such as databases or neural networks, with the crucial requirement being its ability to perform relevance search queries. These queries extract the contextual meaning associated with data information, identifying similarities and returning stored information alongside its relevance value, tailored to the specific relevance requirements.

MaxMind innovates further by expanding the request segment of the MLN into a dedicated functional unit. This unit masterfully transforms user requests into query targets for the MR, leveraging the model's capabilities to conduct multidimensional analysis and extraction. This process scrutinizes the request across various focus areas, including its semantics, spatiotemporal information dimensions, potential knowledge, operations, and files, among others, ensuring a comprehensive understanding of the user's intent.

MaxMind's memory output unit aims to forge a LLM context that is both informative and responsive, that is achieved this by embracing the RAG paradigm, where LLM context is generated through multifaceted queries to the MR. As the RAG context is inherently tied to the MR, past experiences become a valuable resource for MaxMind, enabling it to deliver efficient and relevant services in response to user requests. 
MaxMind addresses the challenge of generating sufficient and coherent memory output within LLM's token size constraint. It achieves this through a precision adjustment mechanism that dynamically scales the precision of relevant memory items, taking into account both their relevance and additional information garnered from the query. This intelligent approach ensures that MaxMind adheres to token size requirements while preserving the relative relationships within the generated context, enabling the system to adaptively fulfill its objectives.

The response unit of MaxMind orchestrates a process where a LLM is invoked to take the former context as the base to perform its reasoning, encompassing the user's request. the LLM's output encompasses knowledge, task planning, and tools necessary for task completion, along with a quantitative assessment of each memory item's contribution to the overall response. The response unit then meticulously establishes execution and validation frameworks, strategizing and executing actions based on its output, invoking tools as needed, and rigorously evaluating subsequent outcomes to ascertain their alignment with user expectations. It further disseminates the results of these executions or tool invocations, empowering users to engage in inspection and adjudication whenever necessary. In scenarios where the output falls short, the response unit send back the contribution data from the LLM's output to the output unit to inversely adjust the precision of utilized memory items, iteratively engaging the LLM and execution evaluation mechanisms to autonomously enhance the task's effectiveness in response to user requests.

At the forefront of MaxMind lies its distinctive Memory Input Unit. This design transforms task experiences into lasting memories. It expertly compiles the execution intricacies of a task into a coherent context, which is then presented to a AI module (may be a LLM) for a definitive judgment on task completion. Irrespective of the outcome—be it success or failure—a concise summary is crafted, serving as the foundation for memory enhancement.

The update unit of MaxMind assumes a important role in refining and preserving memories. It meticulously converts the received memory summaries into structured memory items, adhering to the stringent format requirements of the MR, and securely archives them. Furthermore, this unit dynamically manages the MR, orchestrating the organization, compression, or even the deliberate forgetting of memory items as dictated by the system's needs.

In MaxMind based systems, the typical operation process generally begins with a task request. This request is then converted into a multi-dimensional query object in the Request Encoding unit; the query object reaches the Memory Output unit, which uses a relevance and accuracy correlation search to retrieve memories from the MR and combines them with the user's request to form the context for the main reasoning LLM, and the process moves to the Response Unit; the Response Unit first calls the LLM to generate a solution or tool code, and attempts to execute and provide feedback, improving the inference of the task by returning to the Memory Output unit and modifying the context; after the task is successful or fails, the process enters the Input unit to summarize the task process and generate memories; the process enters the Update Unit to store the new memories into the MR.


\section{Design of a MaxMind Instance} 
We hope to evaluate the impact of integrating MaxMind into the SOTG system, necessitating a robust test suite for complex tasks, which is currently rare and labor-intensive to create. We appreciate the SheetCopilot \cite{li2024sheetcopilot} team's contributions, including an extensive test suite of 221 tasks and open-source code for direct use. We have developed MaxMind4Sheet system, based on the SheetCopilot framework, features an enhanced state machine and MLN capability for rapid validation of MaxMind's key mechanisms.



SheetCopilot abstracts spreadsheet functionalities into virtual APIs, enabling LLM-software interactions. It uses an external knowledge base and real-time feedback to refine LLM-generated solutions. However, SheetCopilot's feedback mechanism is simple and task-independent, maintaining stability but lacking task-based learning. In contrast, our MaxMind4Sheet enhances this with task-specific memory recycling for improved performance and adaptability.

MaxMind4Sheet utilizes PostgreSQL equipped with the PGvector plugin as its MR, enabling accurate memory storage and vector-based memory querying. It establishes multiple keys to capture various attributes of memories, including ID, Task Memories (memories), and Knowledge (knowledge) etc. 
In MaxMind4Sheet, task memories are defined as descriptive narratives of sheet task processes, encompassing requested content and system-generated solutions. 
Three precision levels—"original", "concise", and "brief"—are defined for task memories and knowledge. "Original" includes initial task requests, context interactions, and raw knowledge text. "Concise" focuses on the final task actions and key knowledge points, optimizing token usage. "Brief" retains only the problem and operation name, simplifying memory representation.

For vector embedding of information or memories, MaxMind4Sheet adopts the nomic-embed-text encoder [https://ollama.com/library/nomic-embed-text]. Meanwhile, for searching through information or memories, it employs the Hierarchical Navigable Small World (HNSW) [https://neon.tech/blog/pg-embedding-extension-for-vector-search] method for high-dimensional data search, ensuring efficient and accurate retrieval.

Unlike SheetCopilot, MaxMind4Sheet employs our distinct RAG mechanism to generate a long context for the LLM. It decomposes the request text into two dimensions: Task Memories and Knowledge, facilitated by a lightweight llama3-8B model. The workflow of the three mechanisms—querying, filtering, and long-context generation for memory output—is as follows:
Firstly, based on the decomposed request across these two dimensions, MaxMind4Sheet retrieves the six most relevant memories from its MR: up to three Task Memories and up to three Knowledge items (queries). These are then sorted by relevance, with the corresponding precision levels selected for each. This step involves precision-based information extraction (filtering) and prepends labels like "Key Reference" and "General Reference" to the respective memory items. Together with SheetCopilot's original prompt and feedback information, these form the long context.

MaxMind4Sheet uses llama3.1-70B as its main LLM with a 128K tokens limit. When the context exceeds the threshold, the memory output unit calls "Memory Extraction" modules to downgrade adjustable precision parts and recompose the context until the size fits within the limit.

\begin{figure}[htb]
\centering
\includegraphics[width=0.5\textwidth, trim=0cm 5cm 0.5cm 0cm, clip]{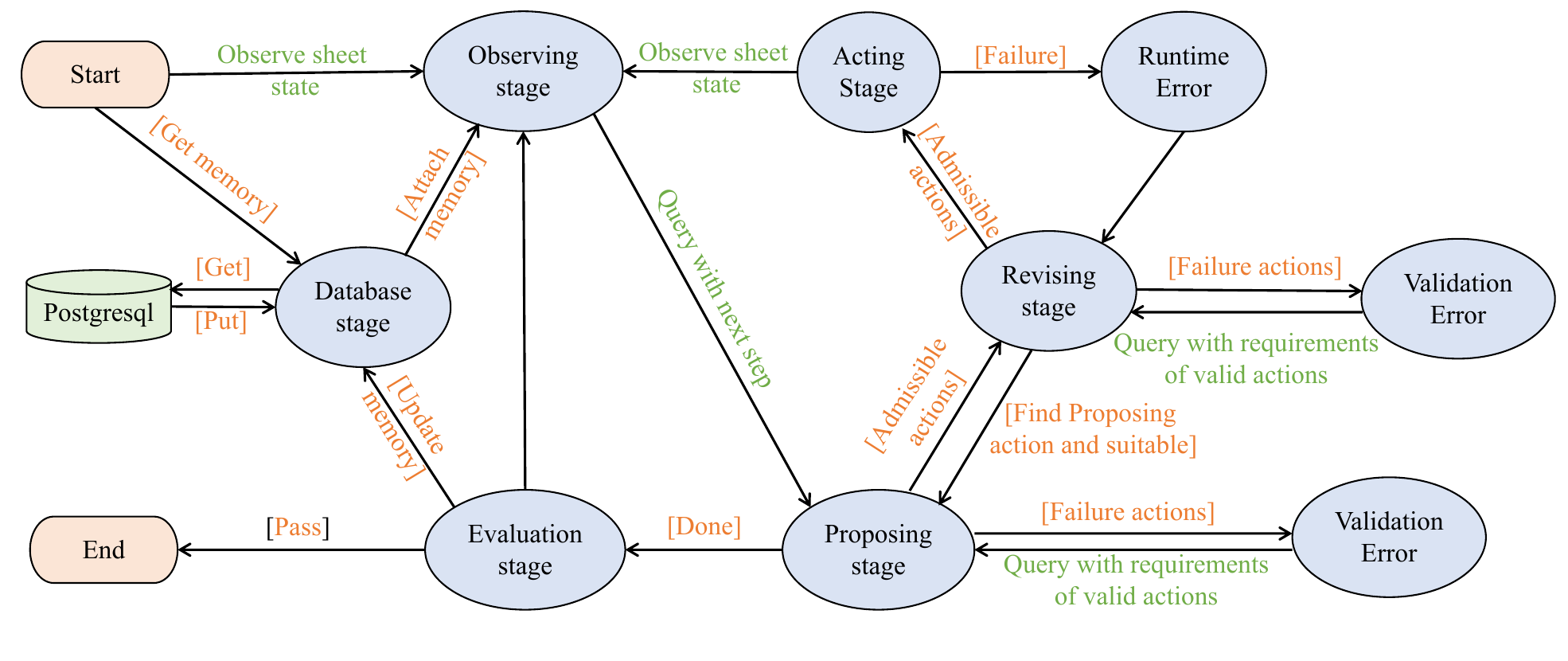}
\caption{State machine for task planning in MaxMind4Sheet}
\label{fig3333}
\end{figure}

Furthermore, MaxMind4Sheet revamps the response feedback mechanism to directly evaluate the execution results, as opposed to SheetCopilot's approach of merely testing EXEC@1 and conducting offline PASS@1 assessments. The evaluated results are then fed back to the Proposal Stage. MaxMind4Sheet's execution state machine has been refined to follow the process outlined in Fig \ref{fig3333}.

Compared to the state machine loop of SheetCopilot, MaxMind introduces two significant enhancements. Firstly, during the Observing Stage, relevant memories and knowledge are extracted from the MR as key references. 
Secondly, we incorporate an evaluation stage, and upon passing the evaluation, the memory is transitioned to a database stage for retention before ultimately reaching the end state.
The state machine's operation involves multiple invocations of LLMs, continuously optimizing the reasoning plan through contextual improvements. This may retrigger memory searches or automatically prompt knowledge updates in the MR (by leveraging online searches for related issues). MaxMind has devised a feedback-based contextual improvement strategy encompassing two aspects: Firstly, updated prompt words are generated based on the current state and evaluation of the state machine. Secondly, each LLM invocation requires the model to assess the contribution of memories attached to the context in its reasoning output, categorizing them into five different contribution levels. These levels are then used to calculate new precision values, adjusting the precision of additional knowledge in the context through feedback.

Upon reaching the End or Fail state, the task's success is evaluated, and a corresponding label is assigned. Following this, the Memory Input Unit collects process information, including task requests, decomposed multi-dimensional query objects, LLM contexts and outputs during state machine operation, logs from the action module in the response unit, evaluation module results, and the final success status. This information is then used for memory summarization.

The Memory Update Unit summarizes the current session content by different precisions and types, calculates embeddings for each, and stores the summarized memories and embeddings in the PostgreSQL database. The two types of memories are automatically generated by the large model llama3-8B based on segmented human-machine interaction transactions or automatic organization of original memories. Each memory transaction can be converted into one or more of these two memory types, and multiple memory items of the same type may correspond to a single interaction transaction.

\section{Experiments}
\subsection{Metrics and Environment of Experiments}
We employ a test set comprising 221 spreadsheet tasks from SheetCopilot and utilize state-of-the-art Large Language Models (LLMs) to simultaneously evaluate both SheetCopilot and MaxMind4Sheet. The objectives of this evaluation encompass: verifying the reasoning performance of various LLMs when provided with a memory reference; and validating the effectiveness of memory cycling, precise memory extraction, and memory feedback mechanisms within MaxMind by comparing its outcomes with those of SheetCopilot on the same dataset.

For the sake of facilitating comparison, we adopt the same evaluation metrics as SheetCopilot\cite{li2024sheetcopilot}, which include: (1) Exec@1, the proportion of solutions provided by the system that can be successfully executed; (2) Pass@1, the proportion of generated spreadsheet states that fully satisfy the task requirements, based on the system's proposed solutions; (3) A50 and A90, computed as the 50th and 90th percentiles, respectively, of the ratios of atomic operations in the system-generated solutions to the number of operations in the ground truth, across all tasks.
Due to the high expense of GPT- series (the primary LLM used in SheetCopilot), we adopted the open-source and free llama3.1-70B as our primary LLM. Additionally, for some comparative tests, we also utilized llama3-70B, llama3-405B, and GPT4o, respectively.

Hardware\&Software: Intel(R) Xeon(R) Platinum 8358 CPU and 8 NVIDIA A800-SXM4-80GB GPUs, Ubuntu 18.04 server and Windows 10 Desktop, Office 2016.


\subsection{Recalibration of Reference Frame}
The experimental compares the performance of our system with SheetCopilot using 221 test tasks from the SheetCopilot project. We first calibrate the performance of SheetCopilot (using its GitHub source code) because differences in our LLM and testing environment might affect results. Table \ref{tab-SC} shows calibrated SheetCopilot performance on four recent LLMs, revealing notable differences from the original paper (average Pass@1 was about 40\%). To ensure these discrepancies aren't due to inappropriate configuration, we tested the core performance mechanisms mentioned in the original paper, as seen in Table \ref{tab-SC-four}, confirming all factors contribute positively. Thus, differing test environments significantly impact results, so we use Table \ref{tab-SC} as a baseline for MaxMind4Sheet performance tests.

\begin{table}[h]
\caption{Recalibration of Performance for SheetCopilot on State-of-the-Art LLMs (red marked data serve as our tests' baseline)}%
\footnotesize
\tabcolsep 17pt 
\begin{tabular*}{0.48\textwidth}{ll|ll}
\toprule
Model     &Dataset    & Exec@1    & Pass@1   \\ \midrule
GPT3.5    &  \multirow{4}{*}{100\%}      & 94.12\% & 16.29\%\\
GPT4o     &        & 87.78\% & 22.17\%\\
llama3.1:70b  &        &{\color[HTML]{FE0000}91.78\%}  &{\color[HTML]{FE0000}21.46\%} \\
llama3.1:405b &        & 89.14\% & 19.00\%   \\ \bottomrule
\end{tabular*}
\label{tab-SC}
\end{table}

\begin{table}[]
\caption{Contributions of Core Mechanism in SheetCopilot
}
\label{tab1}
\resizebox{\columnwidth}{!}{%
\begin{tabular}{@{}l|llll|lll@{}}
\toprule
No. & State      & Error           & doc.       & examples   & Exec@1  & Pass@1  \\ \midrule
1   &  —      & —          & —     & —    & 0\%   & 0\%   \\
2   & —     & —          & \checkmark & \checkmark & 21.27\% & 11.31\% \\
3   & \checkmark & —          & \checkmark & \checkmark &22.62\% &12.22\%\\
4   & —     & \checkmark      & \checkmark & \checkmark &86.88\%&20.81\%\\ \midrule
5   & \checkmark & \checkmark      & —     &  —    & 9.87\%& 1.35\%\\
6   & \checkmark & \checkmark      & \checkmark &  —    &57.01\%& 9.05\%\\
7   & \checkmark & \checkmark      & \checkmark & \checkmark &91.78\%&21.46\%\\ \bottomrule
\end{tabular}}
\label{tab-SC-four}
\end{table}

\subsection{Performance Evaluation}
It must be acknowledged that the amount, type, and quality of memories stored in the MR of the system will all have an impact on the accuracy of subsequent problem-solving. Therefore, an absolute conclusion on the impact of memory on performance cannot be given. This paper adopts an alternative approach to demonstrate the impact of MaxMind4Sheet's memory loop on subsequent tasks. Specifically, we first allow the system to run through a complete test set and save the successfully completed tasks as memory items. Then, we rerun the test set with the system equipped with these memories to observe the completion of tasks in this round of testing. 

As shown in Table \ref{tab-perf}, using Llama3.1-70B as the primary inference model, the first row of data represents the baseline achieved by SheetCopilot, the second row displays the performance test results of MaxMind4Sheet running 221 tasks in the first round, and the third row presents the test results of the second round after retaining the memory from the first round. It can be observed that MaxMind4Sheet's Pass@1 gradually improves significantly over the two rounds of testing (3.3\% increase in round 1, 6.1\% increase in round 2), indicating that the memory acquired from preceding tasks plays a role in the correct execution of subsequent tasks, which aligns with the expectation that success rates incrementally increase with the accumulation of experience. When focusing on the Exec@1 metric, we find that SheetCopilot achieves the best performance in this aspect, while MaxMind4Sheet's performance decreases as the number of rounds and memory increase. We believe this is largely related to the current design of the memory mechanism. The current version of MaxMind4Sheet only retains the memory of tasks executed correctly and lacks a reasonable design for the memory mechanism of tasks that can be executed but with incorrect results. This leads LLM to prioritize the correctness of task results during inference, partially neglecting executability, which results in a decrease in executability rates. We also pay attention to the A50 and A90 test metrics. Here, it can be observed that with the incorporation of successful task memories, MaxMind4Sheet's metrics in the first round of testing initially increase relative to the baseline but begin to decrease in the second round. This can be explained by the fact that when there is not much memory initially, the system can only retrieve memories with relatively low relevance, which introduces relatively more interference to inference. However, as memory increases, the relevance of the memories referenced by each task continues to improve, reducing interference, and thus, A50 and A90 start to decrease. Although this test demonstrates that the introduction of memory loops can effectively improve the task accuracy of the SOTG system, there are still many complex issues to be studied and resolved in designing the mechanisms for retaining and retrieving task memories.

\begin{table}[]
\caption{Performance Evaluation}
\resizebox{\columnwidth}{!}{%
\begin{tabular}{@{}ll|llllll@{}}
\toprule
System     &Dataset   & Exec@1    & Pass@1    & A50  & A90  \\ \midrule
SheetCopilot   &  \multirow{3}{*}{100\%}    & 91.78\% & 21.46\% & 1.5 & 3.0 \\ 
MaxMind4Sheet-r1   &      & 76.02\% & 22.17\% &  2.0&  3.6  \\ 
MaxMind4Sheet-r2   &      & 72.40\% & 23.53\% & 1.5  & 2.9  \\ \bottomrule
\end{tabular}
}
\label{tab-perf}
\end{table}

\subsection{Memory Transfer Capability and System Efficiency}
We hope to verify that the MaxMind4Sheet system support the transfer and sharing of memories and experiences. 
The corresponding experimental design is as follows: Firstly, we use GPT4o and Llama3.1-405B as the main LLM respectively, allowing MaxMind4Sheet to undergo two rounds of testing on a task set consisting of 221 tasks, accumulating MR. Then, we transfer the MR to MaxMind4Sheet with the main LLM switched to Llama3.1-70B, and observe the performance in the following testing. 
At the same time, we also hope to verify the contribution of examples and introduced memories to task accuracy.
The experimental results are shown in Table \ref{tab-tran}. The first row of data is the test result without relying on any memories but with examples added (which is actually equivalent to SheetCopilot); the second row of data is the test result relying on memories but without adding examples; the third row is the test result integrating memories and examples support. Firstly, it can be seen that the transferred memories have significantly improved the correctness of task completion based on Llama3.1-70B. This is because different LLMs have significant differences in tasks that produce correct results during reasoning. Therefore, transferring a combination of multiple successful memories can significantly improve the correctness of system task execution. 
The results also show that replacing examples with memory improves execution efficiency by 25\% compared to SheetCopilot, while not significantly affecting the pass rate of task solving.
\begin{table}[]
\caption{Efficiency and Mem-Transfer (MaxMind4Sheet)}
\resizebox{\columnwidth}{!}{%
\begin{tabular}{@{}ll|ll|llll@{}}
\toprule
Model     &Dataset &example     &Memory     & Exec@1    & Pass@1     &Time     \\ \midrule
\multirow{3}{*}{llama3.1:70b}   &\multirow{3}{*}{15\%}  &\checkmark  &—           & 83.87\%  & 16.13\%    &42.9min  \\
    &  &—            &\checkmark & 80.65\%   & 54.06\%    &\textbf{36.0min}    \\ 
    &  &\checkmark  &\checkmark & 80.65\%   & 58.06\%    &48.1min  \\ \bottomrule
\end{tabular}}
\label{tab-tran}
\end{table}

\subsection{Evaluation of Memory Precision on Task Performance}
To evaluate the influence of memory precision on performance and to bolster our information value-based RAG approach, we tested the impact of different memory precisions on MaxMind4Sheet's Exec@1 and Pass@1 metrics, building upon the previous test methodologies for memory transfer capability. We separately employed three precisions (without selecting memories based on relevance as done in this test) to realize RAG, supporting the reasoning of Llama3.1-70B. As shown in Table \ref{tab-preci}, it is evident that both Exec@1 and Pass@1 results are significantly affected, with concise precision yielding the most favorable pass rate, original precision coming second, and brief precision proving to be the least effective. These findings provide preliminary validation and insights, suggesting that higher-precision memories offer more assistance compared to lower-precision ones; however, exceedingly high precision might introduce noise during LLM inference, potentially causing interference.

\begin{table}[]
\caption{Evaluation of Memory Precision on Task Performance for MaxMind4Sheet}
\resizebox{\columnwidth}{!}{%
\begin{tabular}{@{}lll|lllll@{}}
\toprule
Model     &Dataset   &Precision & Exec@1    & Pass@1    & A50  & A90  \\ \midrule
\multirow{3}{*}{llama3.1:70B}   &  \multirow{3}{*}{15\%} &original   & 80.65\% & 58.06\% & 1.6  & 2.8  \\ 
   &   &concise   & 93.55\% & 61.29\% & 1.5&  2.5  \\ 
   &   &brief   & 80.65\% & 32.26\% & 1.4  & 2.7  \\ \bottomrule
\end{tabular}
}
\label{tab-preci}
\end{table}

\section*{Conclusion and Future Work}
An SOTG system built on LLMs requires learning from experience to handle complex problems. We introduce the Memory-Loop Network concept, inspired by human intelligence, allowing continuous accumulation and referencing of system experiences. Our LLM-based MLN model, MaxMind, incorporates a memory loop for interaction with LLMs and an adaptive RAG method that optimizes experience selection based on knowledge value, enhancing task handling. We validate this model using the SheetCopilot system, creating MaxMind4Sheet and testing it on 221 Excel tasks. Experiments show a 3\%-6\% performance boost per round and an up to 25\% increase in efficiency with memory recycling. MaxMind's task memories are transferable, addressing retraining challenges for specialized tasks, thus enhancing LLM capabilities in software automation and tool generation.

MaxMind4Sheet is only a simple verification of the MaxMind model concept, rather than a full implementation. There are still many knowledge gaps in how to summarize and generate memory, as well as its reasonable and efficient use, which require further research and extensive experimentation. Additionally, there is a severe lack of datasets and scenarios in SOTG, which we will gradually explore in our future work.


\end{document}